\documentclass[aps,prl,twocolumn,superscriptaddress,showpacs,english]{revtex4-1}

\usepackage[T1]{fontenc}
\usepackage[latin9]{inputenc}
\usepackage{babel}
\usepackage{amsmath}
\usepackage{amssymb}
\usepackage{wasysym}
\usepackage{graphicx}
\usepackage{xcolor}
\usepackage{bm}          

\definecolor{mydarkgreen}{rgb}{0.0,0.5,0.0}
\definecolor{friebrick}{rgb}{0.698,0.1333,0.1333}
\usepackage[linktocpage=true,
  colorlinks=true, 
  pdfborder={0 0 0},
  linkcolor=blue,
  citecolor=friebrick,
  filecolor=yellow,
  urlcolor=mydarkgreen,
  bookmarks,
  pdfauthor={},
]{hyperref} 

\newcommand{\tc}{T$_{\text{c}}$}

\newcommand{\fcc}{$Fm\overline{3}m$}

\newcommand{\Roma}{Dipartimento di Fisica, Universit\`a di Roma La Sapienza, Piazzale Aldo Moro 5, I-00185 Roma, Italy}
\newcommand{\UniTokyo}{Department of Applied Physics, University of Tokyo, Tokyo 113-8656, Japan}
\newcommand{\Tokyo}{RIKEN Center for Emergent Matter Science, 2-1 Hirosawa, Wako, 351-0198, Japan}
\newcommand{\Jilin}{State Key Laboratory of Superhard Materials, College of Physics, Jilin University, Changchun 130012, China}
\newcommand{\ICFS}{International Center of Future Science, Jilin University, Changchun 130012, China}

\newcommand{\Mainz}{Max-Planck Institute for Chemistry, Hahn-Meitner-Weg 1 55128 Mainz, Germany}
\newcommand{\Sendai}{Department of Physics, Tohoku University, Miyagi 980-8578, Japan}

\begin{document}

\title{Reaching room temperature superconductivity by optimizing doping in LaH$_{10}$?} 

\author{Jos\'e~A. Flores-Livas} \affiliation{\Roma}\affiliation{\Tokyo} 
\author{Tianchun Wang}          \affiliation{\UniTokyo}
\author{Takuya Nomoto}          \affiliation{\UniTokyo}
\author{Takashi Koretsune}      \affiliation{\Sendai}
\author{Yanming Ma}             \affiliation{\Jilin}   \affiliation{\ICFS}
\author{Ryotaro Arita}          \affiliation{\UniTokyo}\affiliation{\Tokyo}
\author{Mikhail Eremets}        \affiliation{\Mainz}

\date{\today}
\begin{abstract}
Intuitively, doping represents one of the most promising avenues for optimization of best prospect superconductors (SC) such as conventional high-pressure SCs with record critical temperatures. However, doping at high pressure (HP) is very challenging, and there is not a proved route to achieve it in a controlled fashion. Aided by computing simulations, we show that it may be plausible to start by alloying primary materials and subsequently incorporate high ratios of hydrogen at moderates pressures ($\approx$1.5 Mbar). Our theoretical results evidence the possibility to tune the electronic structure of LaH$_{10}$, increase the density of states at the Fermi level by doping of various elements and hence change their superconducting properties. We found aluminium to increase the occupation at the Fermi level by more than 30\,\%. Incorporation of other elements such as Si, Ge, H, Ir, Ca, and others with a varying percentage also play in favour to tune the electronic structure. More importantly, these predictions lie in experimentally attainable doping levels. Also, for the first time, we shed light on how the formation of defects and vacancies influence on the electronic structure of a HP-hydride superconductor. The concepts presented in this work can be extended to other high-pressure, hydrogen-based superconductors such as H$_3$S. Arguably, doping is one of the promising paths to reach room-temperature superconductivity, a Holy grail of condensed matter physics.  
\end{abstract}

\maketitle


With the advent of hydrides as high-temperature superconductors, particular emphasis has been put in studying systematically binaries and ternaries to find materials higher transition temperatures~\cite{Zurek_review,Pickard_review,Flores_review}. 
While there are particular predictions that suggest that some binaries such as CaH$_6$~\cite{CaH6_wang2012superconductive,PRL_clathrate_Ma-2017}, with predicted transition temperature (\tc) of 230\,K or ternaries as the electron-doped Li$_2$MgH$_{16}$~\cite{PRL_Ma_LiMgH_2019,Li2MgH16_Jun-Hyung_2020} that may show a \tc\ well above 300\,K, their complicated stoichiometry, pressure ($>$~2.5\,Mbar), or other factors make these systems elusive to synthesis~\cite{hot_Ma_view_point_PRL2019}.  

An alternative solution to reach the complex synthesis and that has been overlooked experimentally is doping. 
It is well known that by introducing sufficient electron- or hole-donating impurities, one can transform semiconducting systems into metallic phases.  This strategy has been successfully demonstrated by inducing superconductivity in several semiconductors (see Ref.~\cite{bustarret2015superconductivity} for a recent review). For instance, doping in carbon allotropes are predicted to be superconductors at zero  pressure~\cite{superconductivity_Tcarbon,lian2017phonon,hydrogenated_nanostructures}. 
Recent predictions points the importance on hydrogenic bands to enhance \tc\ in known superconductors such as MgB$_2$~\cite{PRL_2019_H-MgB2}. 
Similarly, the idea of doping under pressure has been used to predict that 
insulating systems such as H$_2$O~\cite{H2O} and polyethylene~\cite{polyethylene} will transform into metallic phases 
and display BCS superconductivity. 
Other works have focus on the ternary H-S-C~\cite{Ying_H3S_CH4,Cui_H3S_CH4} phase, 
doping~\cite{PRB_boeri_inflluence,H3SP_doped_PRB2016}, and found the potential to 
increase \tc\ for the parental phase (H$_3$S). While these theoretical premises are technically sound, 
experimentally there is no precise route to reach such doping levels in a controlled way that help to corroborate these predictions~\cite{struzhkin_2020_remaining}. 

In this work, we screen for the best elements that can dope LaH$_{10}$, 
currently the {\it best} superconductor with an extraordinarily high-\tc\ of 250\,K. 
The ensemble of results issued of first principles shows that it is possible to tune the electronic properties of LaH$_{10}$ and induce a sizable change in the electronic occupation at Fermi level. 
The interplay between structure and dopant (electronic behaviour) is discussed, and essential hints are proposed. 
Finally, to connect our predictions with experimentation, in this first research article we propose a simple, yet clear idea that stands an excellent chance to be generalized in the case of hydrides materials and other materials subject of pressure studies. 

Based on the conclusions found by Errea et al.~\cite{Errea_Nature_2020} and in Ref.~\cite{Flores_review}. 
We propose the following series of statements as recipe guidance for our study. 
To maximize \tc\ in hydrides or other light-mass superconductors under pressure, 
\begin{itemize}
    \item It is desirable to opt for highly symmetric cages of hydrogen.
    \item To have a compact (dense) basis of lanthanum positions (or guest element) that minimize enthalpy. 
    \item The electronic density of state should ideally show a Van Hove singularity (VHS) pinning at or nearby the Fermi level. 
    \item The heat of formation of solid or liquid metal alloy for the guest atoms should be favourable, i.e. the dopant atom should be miscible in the guest matrix, i.e. X atom miscible to La, in the case of LaH$_{10}$. 
\end{itemize}
Finding the best element that holds the above criteria is something that can be accessed using state-of-the-art first principles, in return, there is a large computational overhead (see details in Ref.\,\footnote{Energy, enthalpy, atomic forces and stresses were evaluated at the DFT level with the GGA-PBE parametrization to the exchange-correlation functional. 
A plane wave basis-set with a cutoff energy of 650\,eV was used to expand the wave-function together with the projector augmented wave (PAW) method as implemented in the Vienna Ab Initio Simulation Package~{\sc vasp}~\cite{VASP_Kresse,GPU_VASP_2012}. 
To simulate doping substitutions, different supercell sizes were created using the $Fm\overline{3}m$ phase of LaH$_{10}$ at 150\,GPa. The valence configuration for La was treated as 5$s^2$5$p^6$5$d^1$6$s^2$,
H as $1s^1$ and the rest of elements the POTCAR prepared for high pressure or with the highest number of valence electrons was used. 
Three different percentage of doping-substitution were considered in this study. 
For 12.5~\% doping, a supercell of 2$\times$2$\times$2 (8~f.u. of LaH$_{10}$) with 88 atoms was used. 
The sampling of the Brillouin zone (BZ) was performed with a $\Gamma$-6$\times$6$\times$6 $k$-mesh. 
For 3.70\%, a supercell of 3$\times$3$\times$3 (27~f.u. of LaH$_{10}$) with 297 atoms was employed. 
The integration of the Brillouin zone was carried out with a $\Gamma$-4$\times$4$\times$4 $k$-mesh. 
Finally, the 1.56\% doping was simulated with a supercell of 4$\times$4$\times$4 (64~f.u. of LaH$_{10}$) 
with 704 atoms. The integration of the Brillouin zone was carried out with a $\Gamma$-3$\times$3$\times$3 $k$-mesh. 
For each of the size supercell, independently of chemistry and logic reasons of substitutability, 
a grand total of 70 elements were tested (omitting 13 lanthanides, except for Lu). 
A two-level geometry relaxations were carried out: the first one involved only volume optimization with fix La-basis and the second with breaking symmetry in the dope substitution site. 
Different local relaxations were carried out per element per cell (aggressive and smooth) and accelerated using the GPU ported version of {\sc vasp}~\cite{GPU_VASP_2012} which allowed us to cut the computational overhead by a factor of fourth or more. 
Tight criteria on forces, less than 2\,meV/\AA\ and the stresses less than 0.1~eV/\AA$^3$ were used as a control to judge convergence. 
The electronic density of states were integrated using tetrahedron method for a single formula unit cell with 11 atoms, $\Gamma$-44$\times$44$\times$44 (2,168 points in of the irreducible part of the BZ). 
For supercell calculations, we opt for Methfessel-Paxton with the following 
dense $k$-meshes: 
for cells with 88 atoms, $\Gamma$-24$\times$24$\times$24 (413 points), 
for cells with 297 atoms, $\Gamma$-16$\times$16$\times$16 (145 points), and 
for cells with 704 atoms, $\Gamma$-10$\times$10$\times$10 (47 points). }).  Still, it costs can be minimized using smart high-throughput with autonomous computing tools (see details in Ref.~\cite{Flores_review,flores2020crystal}, on where similar tools have been deployed). 

\begin{figure}[t!]
\includegraphics[width=1\columnwidth,angle=0]{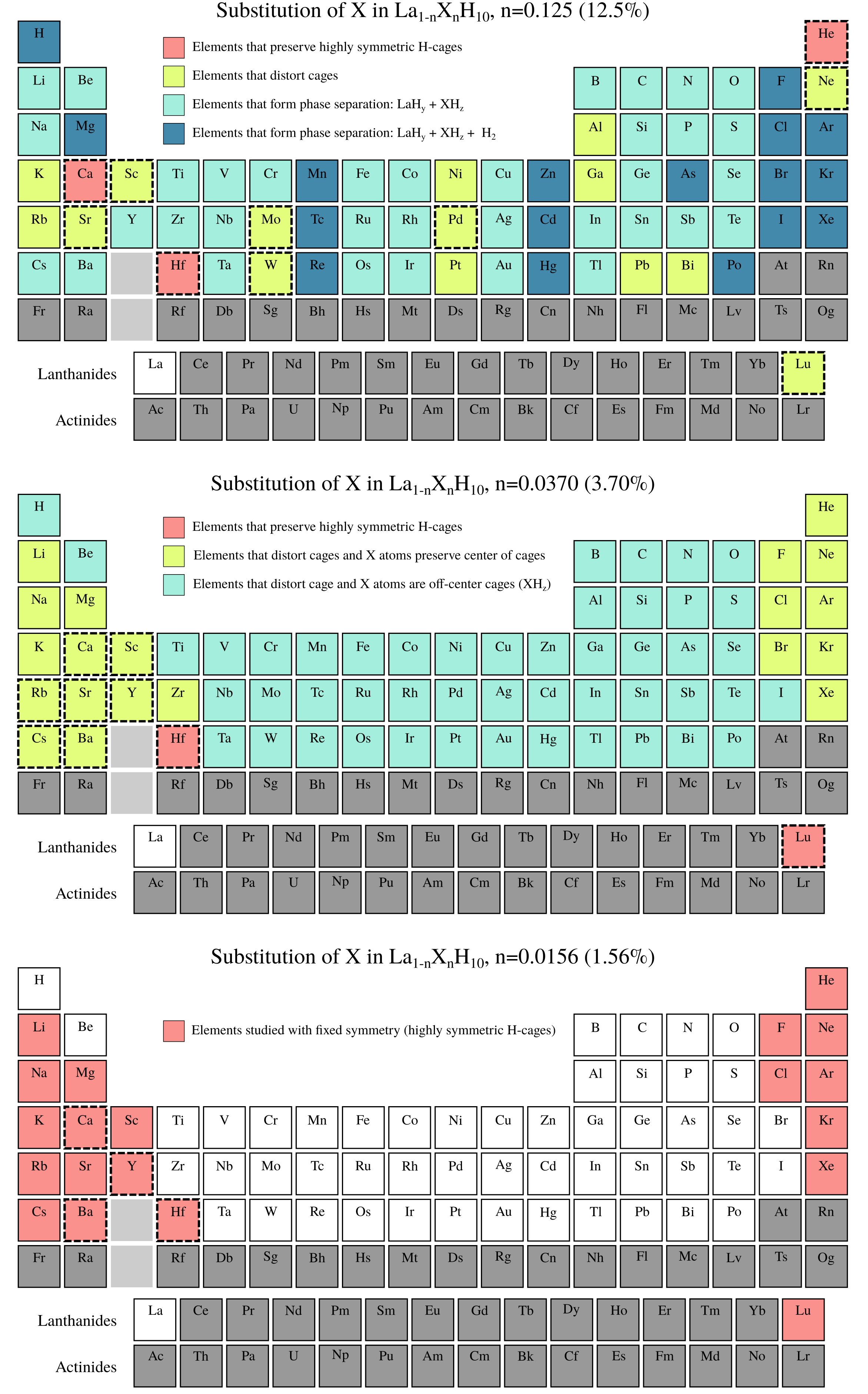}
\caption{~Periodic charts showing the elements explored as doping-substitution in LaH$_{10}$. 
The pressure is 150\,GPa for the entire study. 
For each element, the color code gives the behavior when substituted (see SM for details). 
At a high percentage, most of the elements destabilized the clathrate. 
At low doping, many elements preserve the H-cages, and some of them show unusual electronic behavior. 
On dashed-squares are elements for which DOS is plotted in Fig.~\ref{fig:DOS}.}
 \label{fig:chart}
\end{figure}

Fig.~\ref{fig:chart} shows three periodic charts, each for the three dopings (12.5\,\%, 3.70\,\% and 1.56\,\%) and the elements considered as substitution on La-matrix. Elements coloured in dark-grey were not considered in this study, due to its reduced availability or complex physics, most of the $4f$-electron valence were excluded except for lutetium. 
Interestingly, with 12.5\,\%, there exist only three elements (colored in red) that match perfectly in the H-cages and leaves them intact, thus preserving a high symmetric pattern. In yellow, all the 15 elements that induce distortions in the hydrogen cages, for comparison these are sizable distortions as the ones in LaH$_{10}$ (monoclinic $C2$ phase) but preserve a guest-host clathrate chemistry. 
Coloured in light-blue, 35 elements, and dark-blue 17 elements, that play in favour of phase separation. 
The analysis based on bonding, XRD patterns and nearest-neighbours, elucidate two levels of phase separation: 
the first one favours (LaH$_{x}$ + XH$_{x}$) and the second one a ternary phase separation, (LaH$_{x}$ + XH$_{x}$) and H$_2$ units detached. 
Details of the structures for these classifications (colour code in Fig.~\ref{fig:chart}) are included in the Supplemental Materials (SM). 

For the entire set of elements, we did explore a higher doping percentage, 25\,\%. 
However, we notice that the physics of the La-H$_{10}$ system change enormously. 
Indeed, for many elements, new crystalline structures are found after a full geometry optimization. 
At higher doping, i.e. 1 atom every 4 La, enthalpy plays in favour of other more stable structures; making the study of the 25\,\% 
to widely deviate from our initial assumptions. To be sure of having meaningful results at such a large percentage, it would be necessary to calculate the formation enthalpy, which will be cumbersome and computationally intractable with such vast channels of decomposition. 
Based on the structural patterns and results from the high-throughput, we claim that our results are valid for small doping percentage, 
approximately below 12\%, where the enthalpy of the parental phase dominates. 

For 3.70\% doping, it is unusual to find only two elements (coloured in red) Hf and Lu that after a full optimization 
(starting with breaking symmetry) relaxed back to the lock position, preserving the highly symmetrical pattern of H-cages. 
In yellow, the elements mostly group I and II, halogen and noble gases that retain symmetric-centres (of fcc La basis) 
but induce deformation of the hydrogenic cages. In light blue, transition metals and block-p elements for which the local environment of the dopant produces XH$_x$ phase (dopant atom off-centred of the fcc basis). 
Based on the results of 3.70~\%, as input for 1.56~\%, we optimized the volume only for particular cases. 
Coloured in red all the elements considered at the centre of cages with highly symmetric hydrogenic cages (no breaking symmetry).  

\begin{figure}[t!]
\includegraphics[width=1\columnwidth,angle=0]{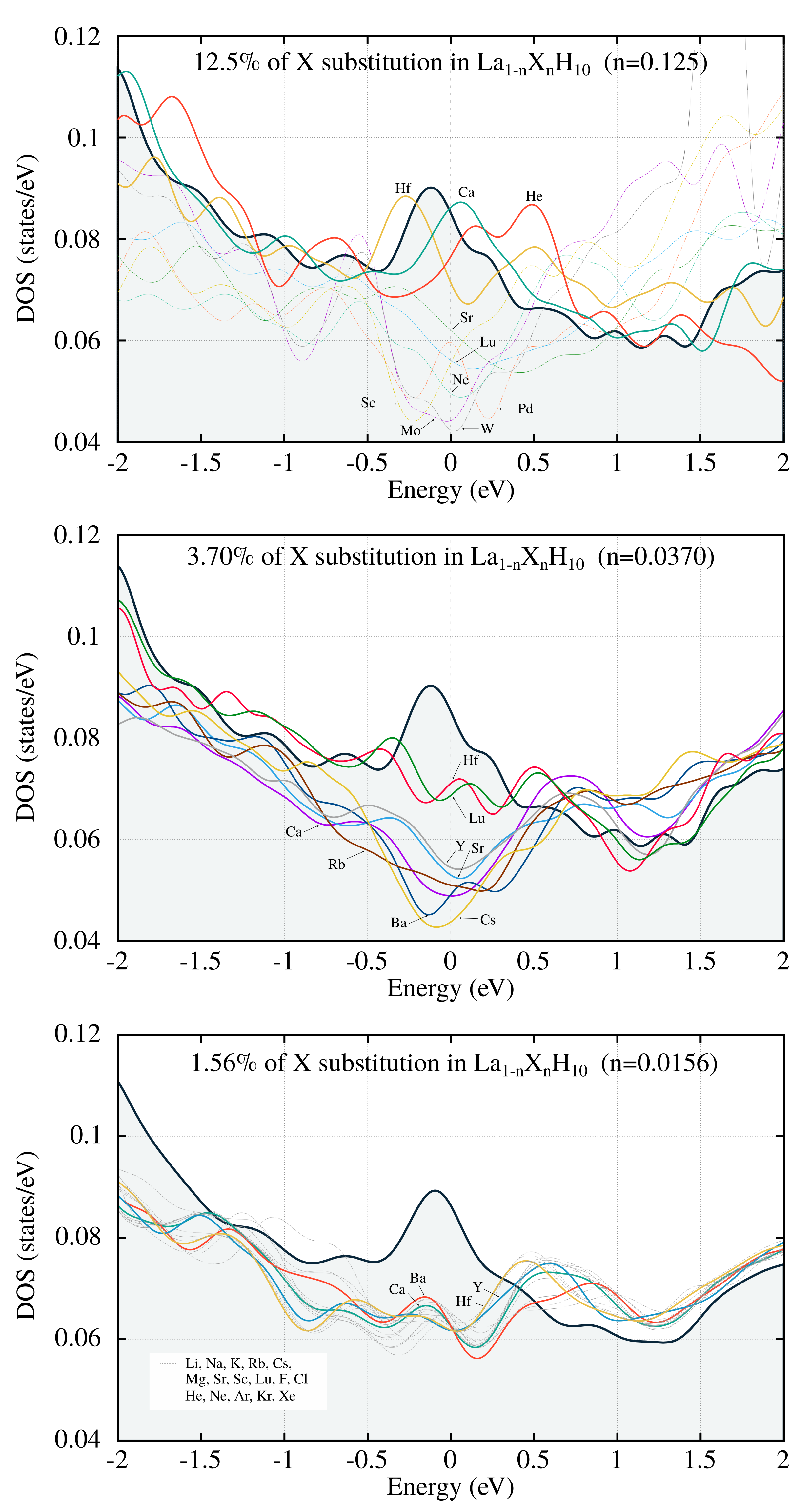}
\caption{~Electronic density of states near the Fermi level for different doping cases (fully optimized geometry). 
The pressure is 150\,GPa for the entire study. In solid lines interesting dopants are shown and lighter-dashed colors cases of less importance.
For reference, the DOS of LaH$_{10}$ is shown in black-line (grey area) calculated in the same supercell.}
 \label{fig:DOS}
\end{figure}

Form the electronic perspective; ideally, one should look for a dopant that increases the density of states at the Fermi level.  
Fig.~\ref{fig:DOS} shows the corresponding density of states (DOS) for selected dopants. 
At 12.5\% doping, the most compelling cases are shown with a solid-line (Ca, Hf and He) and compared to the undoped case of LaH$_{10}$. 
The shape of the DOS is preserved for these cases, Hf shifts in the opposite direction the VHS and Ca, interestingly, 
raises the VHS near the Fermi level. 
At 3.70\% doping, Hf and Lu are the only elements that appear to keep a highly symmetric structure. 
Their DOS however seems to play detrimental to the formation of the VHS. The rest of the elements completely destroys the VHS. 
For the case of 1.5\% doping, the atoms that show a relevant DOS are: Ba, Ca, Y, and Hf. 
However, it is noticeable that the small volume effect of the dopant and the charge transfer within the H-cage and the dopant downturns the VHS. 
It is a robust result to confirm that even low dopings, 3.7\%, induce sizable changes on the shape near the Fermi level of the  LaH$_{10}$ matrix. For the pressure considered in this study (150\,GPa) even at the dilute limit $\sim$1.5\% doping, most of the elements produce a vanish of the electronic singularity. Recently, Akashi~\cite{PRB_VHA_Akashi} has proposed a mechanism for peaking of the density of states in three-dimensional crystals and contrasted for the case of H$_3$S under pressure. Indeed singularities or divergences at DOS are not generally present in the three-dimensional case. What is more appealing, as found by Akashi in H$_3$S, and confirmed in this work for LaH$_{10}$ is the negligible pressure dependence of the VHS (see SM). 
The range of {\it stability} for the VHS lies on a broad scale, in H$_3$S is from 120\,GPa to 240\,GPa. In the case of LaH$_{10}$, it starts from 110\,GPa and extends beyond 300\,GPa. This evidence of the unique electronic features of hydrides and challenges researchers to find other materials, for which the singularity can be stabilized at much low pressure. 

It is clear that the crystalline structure of LaH$_{10}$ and its electronic features~\cite{multiband_LaH10_PRB_2020} are quite unique, and there is very little marge to improve by doping. We can rule out that the following elements, Pt, Ga, Au, B, N, and C, suggested by Grockowiak et al.~\cite{grockowiak2020hot} in a recent work form a higher-order hydride. These elements are likely not forming any ordered structure nor enhancing the superconducting properties of LaH$_{10}$, at least not in the doping, pressure and temperature range studied in this work. 

\begin{figure}[t!]
\includegraphics[width=1\columnwidth,angle=0]{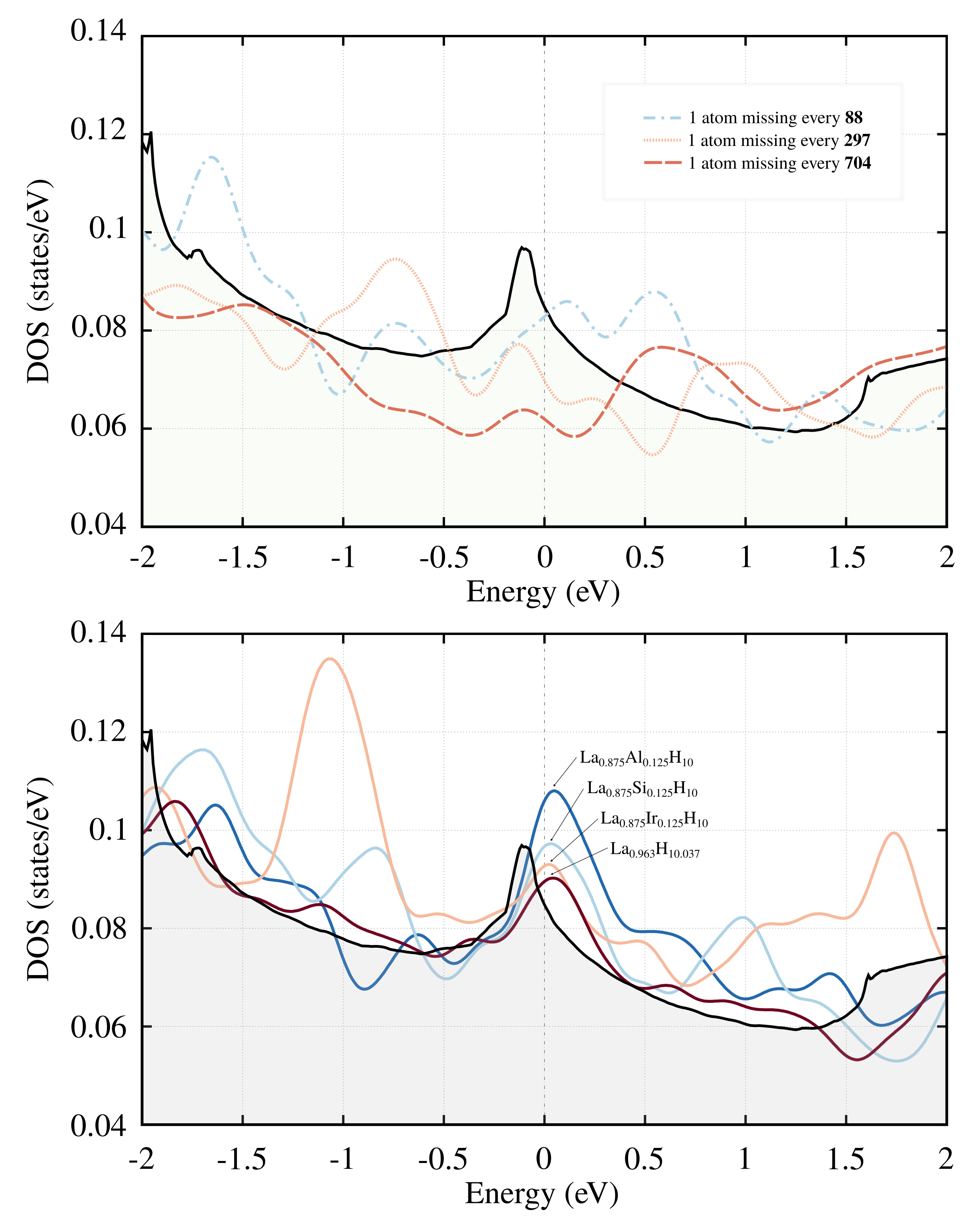}
\caption{~Top panel: DOS comparison of LaH$_{10}$ against a single vacancy in the La-site, for three different concentrations. 
Even 1 atom missing every 704 breaks the VHS.  
Bottom: DOS of the best dopants (assuming that nuclear quantum effects holds highly symmetric H-cages) found in this work. 
In solid-line (with grey area) the DOS of LaH$_{10}$ for a single unit. 
Al, Si, Ir and H, donate the right amount of electrons to the matrix; shifting the VHS and increasing the number 
of electrons at the Fermi level substantially.}
 \label{fig:BEST}
\end{figure}

Moreover, our results are in line with recent evidence by Seho et al.~\cite{LaH10_Jun-Hyung_2020}, 
that highlight the importance of the metal framework of La atoms poses excess electrons at interstitial regions. 
Remarkably, the charge transfer from La to H atoms is mostly driven by the electride property of the La framework, 
where the interaction between La atoms and H cages induces a delocalization of La-5p semi core states to hybridize with H-1s state. 
Thus the bonding nature between La atoms and H cages is characterized as a mixture of ionic and covalent.
Based on our results, we hypothesize that hydrides when successfully synthesized at high pressure, are indeed very special,
and form most of them quasi-perfect crystals. 
We can confirm that in LaH$_{10}$, for instance, even a low concentration defect (in this case a vacancy, 1 atom missing every 700) breaks the VHS, changing the superconductor properties drastically, as seen in the top panel in Fig.~\ref{fig:BEST}. 
We support our claims with the recent evidences by Mikov et al.~\cite{Vasily_D3S_2020} in D$_3$S 
that shows that sizeable \tc's can be measured by performing different annealing procedures~\cite{Vasily_D3S_2020}. 
Similarly, in LaH$_{10}$, two different onset exist, above 260\,K~\cite{Hemley-LaH10_PRL_2019} 
and close to 250\,K with a dome-shape like behaviour under pressure~\cite{Nature_LaH_Eremets_2019}. 
Presumably, in other systems such as yttrium hydride~\cite{Kong_arxiv_YH9_2019,Oganov-Monacelli_YH6} or in thorium hydride~\cite{Semenok_arxiv_ThH10_2019}, a deviation on the measured values of \tc\ seem to correlate more with 
the annealing procedure and less with the the H-content, for a given stoichiometry. 

In the present study, we cannot rule out the possibilities that a complete quantum treatment of nuclear positions (see the work of Errea et al.~\cite{Errea_Nature_2020}) symmetrize for some of the dopants that otherwise distort at the classical level the H-cages. 
To verify this, however, is computationally not tractable in the case of simulating dopants through large supercells 
(hundreds of atoms per primitive cell). 
To rule any potential conflict in our predictions, arising from the lack of full nuclear quantum treatment, 
we cross-check our results by imposing the symmetric structure, relaxing the volume and mapped the DOS for all dopants and different percentages. 
The bottom panel in Fig.~\ref{fig:BEST} shows, for best elements, the expected DOS while preserving a high symmetric pattern. 
Interestingly, Al, Si, Ir and H, not only shift the VHS, but also populate electrons at the Fermi level substantially. 

The Bader analysis~\cite{bader1990chem,tang2009grid} gives the following ionic charges, 
for the undoped-LaH$_{10}$ at 150\,GPa is La$^{1.29+}$, H$_{\rm WP-1}^{0.10-}$, H$_{\rm WP-2}^{0.05-}$.  
For doped cases (assuming quantum effects symmetrize the \fcc\ phases for low doping): 
[La$^{1.30+}$, H$_{\rm  WP-1}^{0.22-}$, H$_{\rm WP-2}^{0.05-}$, Al$^{3+}$],   
[La$^{1.28+}$, H$_{\rm  WP-1}^{0.15-}$, H$_{\rm WP-2}^{0.05-}$, Si$^{1.22+}$], and 
[La$^{1.22+}$, H$_{\rm  WP-1}^{0.05-}$, H$_{\rm WP-2}^{0.07-}$, Ir$^{0.4+}$]. 
The number of electrons at the Fermi level correlates with the excess electrons supplied by the dopant. 
Interestingly, adding one hydrogen has a positive impact on the electronic structure, for 1.5\% 
the ionic charge is [La$^{1.27+}$, H$_{\rm WP-1}^{0.05-}$, H$_{\rm WP-2}^{0.07-}$, H$_{\rm center}^{0.23-}$]. 
However, it should be take with precaution this result; the excess of hydrogen is placed in the centre of the H-cage, 
position that seems odd for a stable structure. Finding a suitable position (interstitial or in the H$_{32}$ framework) will have a beneficial impact by shift the VHS and populating the Fermi level by adding electrons. 

Theoretically, incorporating the right dopants can have significant implications in the electronic properties of LaH$_{10}$. The remaining question is, how do we incorporate these, in a controlled way, without destroying the high symmetry and stoichiometric phase. We propose to start by first, synthesizing the alloy, as for example the 
La$_{0.9}$Al$_{0.1}$ that at low doping limit should follow the ordered fcc of La. 
Then follow the steps of Drozdov et al.~\cite{Nature_LaH_Eremets_2019}: applied pressure 
and incorporate high hydrogen rations and eventually laser-heating followed by cycles of thermal annealing to induce crystallization. 
 
Recently 813 binary alloys were studied and classified using Miedemas theory~\cite{miedema1973simple,miscible_systems_2017informatics}. 
Metallic La is known to be miscible with 15 elements, Al, Tc, Os, Ru, Co, Ir, Rh, N, Pt, Pd, Au, Ag, Cu, Hg, Cd and Zn.
The rest of the elements are a practically uncharted territory and are not knowing to alloy with La. 
Another potential approach to form alloying is to use the valence skipping nature of some elements; 
as it was recently found in GeTe the superconductor-semiconductor-superconductor transition controlled by finely-tuned In doping at high pressure~\cite{PRL_2020_GeTe_doping}. 

In conclusion, this work provides a thorough study of doping and structural stability in LaH$_{10}$ 
for a wide variety of elements (see SM for details).  
The best elements to tune the electronic structure in LaH$_{10}$ at 150\,GPa 
are Al, Si, Ge, Ir, Mg, Ca, Sr, and Ba at 12.5\%. 
Mn, N, Br, Ru and H at 3.7\%. For lower concentrations, the incorporation of dopants
although plays in favour of enthalpy stability, it downturns the stability of the Van Hove singularity. 
For the first time, we shed some light on formation of charge defects and vacancies at high pressure. 
In this work we propose the idea that hydrides when successfully synthesized at HP, 
these can only be formed in almost perfect crystals (free defect) and in perfect stoichiometries. 
Assuming a BCS phonon-mediated superconductor, where \tc\ is proportional to $\omega$exp$(-1/ \lambda)$, where $\omega$ represents the frequency of the phonon mode mediating the electron pairing, and $\lambda$ is the electron-phonon dimensionless parameter representing the pairing strength. Because the electron-phonon coupling is proportional to the DOS at the Fermi level, a peaked DOS is to be a crucial factor in enhancing \tc ~\cite{quan2016van}. Given the enhanced electronic structure, increased occupation of electrons at the Fermi level by >30\,\% with Al, and the ratio of masses, these doping represent a reliable route to further increase \tc\ in LaH$_{10}$ by tens of Kelvins.
However, the current challenge on size supercell, stop of us from performing precise electron-phonon calculations, including anharmonic effects of the doped phases. Instead, we estimated the value of the \tc\ at higher pressure (where anharmonic effects become less critical) at 300\,GPa, for the doped and the pure phase. We found that up to a 15\,\% increase in \tc\ is expected for the Al-doped phase at the harmonic level. 
Experimentally, to verify our predictions, the road to produce these phases is proposed here. First alloying the guest atom. The enthalpy formation of these phases is guaranteed only at low doping concentrations (ideally between 15\,\% to 3.7\,\%). Second to form an alloy-hydride and subsequently pressurize it, for which standard routes have been mastered and are routinely used in high-pressure experiments. 
More importantly, the concepts presented in this work can be extended to other high-pressure, hydrogen-based superconductors such as H$_3$S. Arguably, doping is one of the promising paths to reach room-temperature super-conductivity, a Holy grail of condensed matter physics. 

{\bf Acknowledgements.} 
Without the computational resources provided by the Swiss National Supercomputing Center (CSCS) project s970, this work could not have been possible. We deeply acknowledged these resources. 

\bibliographystyle{apsrev4-1}
\bibliography{main}
 
\end{document}